\documentclass[twocolumn]{aastex631}

\begin{document}

\title{Reconnection nanojets in an erupting solar filament with unprecedented high speeds}

\correspondingauthor{Hui Tian}
\email{huitian@pku.edu.cn}

\author[0000-0002-6641-8034]{Yuhang Gao}
\affiliation{School of Earth and Space Sciences, Peking University, Beijing, 100871, People's Republic of China}
\affiliation{Centre for mathematical Plasma Astrophysics, Department of Mathematics, KU Leuven, Celestijnenlaan 200B bus 2400, B-3001 Leuven, Belgium}

\author[0000-0002-1369-1758]{Hui Tian}
\affiliation{School of Earth and Space Sciences, Peking University, Beijing, 100871, People's Republic of China}
\affiliation{State Key Laboratory of Solar Activity and Space Weather, National Space Science Center, Chinese Academy of Sciences, Beijing, 100190, China}

\author[0000-0003-4052-9462]{David Berghmans}
\affiliation{Solar-Terrestrial Centre of Excellence - SIDC, Royal Observatory of Belgium, Ringlaan -3- Av. Circulaire, 1180 Brussels, Belgium}

\author[0000-0001-9491-699X]{Yadan Duan}
\affiliation{School of Earth and Space Sciences, Peking University, Beijing, 100871, People's Republic of China}

\author[0000-0001-9628-4113]{Tom Van Doorsselaere}
\affiliation{Centre for mathematical Plasma Astrophysics, Department of Mathematics, KU Leuven, Celestijnenlaan 200B bus 2400, B-3001 Leuven, Belgium}

\author[0000-0001-7866-4358]{Hechao Chen}
\affiliation{School of Physics and Astronomy, Yunnan University, Kunming 650500, People's Republic of China}

\author[0000-0002-2265-1803]{Emil Kraaikamp}
\affiliation{Solar-Terrestrial Centre of Excellence - SIDC, Royal Observatory of Belgium, Ringlaan -3- Av. Circulaire, 1180 Brussels, Belgium}



\begin{abstract}

Solar nanojets are small-scale jets generated by component magnetic reconnection, characterized by collimated plasma motion perpendicular to the reconnecting magnetic field lines. As an indicator of nanoflare events, they are believed to play a significant role in coronal heating. Using high-resolution extreme-ultraviolet (EUV) imaging observations from the Extreme Ultraviolet Imager (EUI) onboard the Solar Orbiter mission, we identified 27 nanojets in an erupting filament on September 30, 2024. They are potentially associated with the untwisting of magnetic field lines of the filament. Most nanojets exhibit velocities around 450 km s$^{-1}$, with the fastest reaching approximately 800 km s$^{-1}$, significantly higher than previously reported but comparable to the typical coronal Alfv\'{e}n speed. To our knowledge, these are the highest speeds ever reported for small-scale jets (less than $\sim1$ Mm wide) in the solar atmosphere. Our findings suggest that these nanoflare-type phenomena can be more dynamic than previously recognized and may contribute to the energy release process of solar eruptions and the heating of coronal active regions.

\end{abstract}



\section{Introduction} \label{sec:intro}

The coronal heating problem is one of the most significant and unresolved challenges in solar physics. Among the proposed mechanisms, magnetic reconnection and wave dissipation are considered the two primary processes potentially responsible for heating the solar corona. The most popular magnetic reconnection heating model, initially proposed by \cite{Parker1988}, highlights the importance of magnetic braiding and nanoflares (see the review by \citealp{Klimchuk2017}). Specifically, turbulent convective motions in the lower solar atmosphere can generate braided and twisted magnetic field lines in the corona, particularly within coronal loops. When the misalignment angle between adjacent magnetic field lines becomes large enough, small-scale magnetic reconnection events can occur \citep{Antolin2021,Pagano2021,Cozzo2025}. 
These reconnection processes release stored magnetic free energy, resulting in the heating of the surrounding plasma \citep{Parker1988, Cirtain2013, chenYJ2021, chitta2022}.

One type of direct observational evidence of such reconnection events is known as nanojets, characterized by jet-like features oriented perpendicular to the field lines \citep{Antolin2021,Pagano2021,Sukarmadji2022,Patel2022,Sukarmadji2024,Cozzo2025}. Similar phenomena have also been reported in \cite{ChenHD2017} and \cite{ChenHD2020}. These features are interpreted as a consequence of the slingshot effect caused by the reconnection of curved magnetic field lines experiencing magnetic tension at small misalignment angles. The energy release by nanojets is estimated to be around $10^{24}$ erg, falling within the energy range of nanoflares \citep{Antolin2021}. Their speeds, derived from time-distance maps, typically range between 50 and 300 km s$^{-1}$ \citep{ChenHD2017,ChenHD2020,Antolin2021,Sukarmadji2022,Patel2022,Sukarmadji2024}.
Detection and investigation of nanojets are are hampered by their small spatial scales (approximately 0.5 Mm in width and 1.0 Mm in length) and short lifetimes (lasting no more than 15 s). Therefore, they are more easily identified using data with high spatial and temporal resolutions, such as imaging observations from the Interface Region Imaging Spectrograph (IRIS; \citealp{DePontieu2014}) and Hi-C 2.1 \citep{Rachmeler2019}. 

The Extreme Ultraviolet Imager (EUI; \citealp{Rochus2020}) onboard Solar Orbiter \citep{Muller2020} provides ultra-high resolution imaging of the solar corona in 174 \AA\, channel. In this letter, we report the observation of reconnection nanojets in an erupting filament captured by EUI on September 30th, 2024. These nanojets exhibit unprecedented high speeds, reaching up to $\sim$800 km s$^{-1}$, significantly surpassing all previously reported values. This letter is organized as follows: Section \ref{sec:obs} describes the observation dataset, Section \ref{sec:res} presents our observational results, and Section \ref{sec:diss} provides a discussion and summary of our findings.

\section{Observation}\label{sec:obs}

\renewcommand\tabcolsep{6.pt}
\begin{longtable*}{c c c c c c c c c}
\caption{Properties of observed nanojet events}\label{tab:1} \\
\hline\hline
Jet No. & Time & Duration (s) & Length (Mm) & Width (Mm) & Speed (km/s) & Kinetic Energy (erg) & Direction  \\
\hline
\endfirsthead
\hline
Jet No. & Time & Duration (s) & Length (Mm) & Width (Mm) & Speed (km/s) & Kinetic Energy (erg) & Direction \\
\hline
\endhead
\hline
\multicolumn{8}{r}{Continued on next page} \\
\hline
\endfoot
\hline
\endlastfoot
1 & 23:39:52 & 4 & 0.9 & 0.3 & 449 & $9.23 \times 10^{22}$ & upper right \\
2 & 23:39:56 & 4 & 1.1 & 0.3 & 539 & $2.31 \times 10^{23}$ & upper right \\
3 & 23:40:16 & 2 & 1.0 & 0.4 & 460 & $1.83 \times 10^{23}$ & upper right \\
4 & 23:40:26 & 10 & 0.8 & 0.3 & 156 & $1.44 \times 10^{22}$ & lower left \\
5 & 23:43:37 & 10 & 1.2 & 0.4 & 141 & $2.09 \times 10^{22}$ & lower left\\
6 & 23:43:53 & 8 & 1.6 & 0.5 & 499 & $5.87 \times 10^{23}$ & upper right\\
7 & 23:43:54 & 6 & 1.3 & 0.4 & 478 & $2.68 \times 10^{23}$ & upper right\\
8 & 23:44:02 & 2 & 0.7 & 0.3 & 362 & $7.44 \times 10^{22}$ & upper right\\
9 & 23:44:12 & 4 & 1.2 & 0.3 & 302 & $7.50 \times 10^{22}$ & upper right\\
10 & 23:44:22 & 6 & 1.2 & 0.4 & 172 & $3.30 \times 10^{22}$ & upper right\\
11 & 23:44:32 & 14 & 3.3 & 0.4 & 519 & $1.01 \times 10^{24}$ & upper right\\
12 & 23:44:32 & 4 & 1.0 & 0.3 & 601 & $2.77 \times 10^{23}$ & lower left\\
13 & 23:44:36 & 28 & 6.6 & 0.7 & 770 & $1.14 \times 10^{25}$ & upper right\\
14 & 23:44:38 & 2 & 1.2 & 0.3 & 528 & $1.93 \times 10^{23}$ & lower left\\
15 & 23:44:48 & 16 & 1.9 & 0.5 & 614 & $1.05 \times 10^{24}$ & lower left\\
16 & 23:44:54 & 18 & 3.1 & 0.3 & 513 & $3.54 \times 10^{23}$ & upper right\\
17 & 23:46:00 & 8 & 1.8 & 0.5 & 595 & $1.26 \times 10^{24}$ & lower left\\
18 & 23:46:06 & 4 & 0.6 & 0.5 & 242 & $5.76 \times 10^{22}$ & lower left\\
19 & 23:46:10 & 6 & 0.9 & 0.3 & 323 & $5.77 \times 10^{22}$ & lower left\\
20 & 23:46:20 & 6 & 1.1 & 0.3 & 469 & $1.97 \times 10^{23}$ & lower left\\
21 & 23:46:22 & 12 & 1.2 & 0.6 & 370 & $3.64 \times 10^{23}$ & lower left\\
22 & 23:46:24 & 6 & 1.1 & 0.4 & 128 & $1.57 \times 10^{22}$ & upper right\\
23 & 23:46:28 & 4 & 1.4 & 0.3 & 325 & $1.03 \times 10^{23}$ & lower left\\
24 & 23:46:38 & 2 & 1.2 & 0.4 & 478 & $2.72 \times 10^{23}$ & lower left\\
25 & 23:46:44 & 8 & 1.1 & 0.4 & 464 & $2.44 \times 10^{23}$ & lower left\\
26 & 23:46:46 & 10 & 1.3 & 0.4 & 530 & $3.17 \times 10^{23}$ & upper right\\
27 & 23:46:46 & 14 & 1.7 & 0.5 & 469 & $6.34 \times 10^{23}$ & upper right\\
\end{longtable*}

\begin{figure*}
    \centering
    \includegraphics[width=1.0\textwidth]{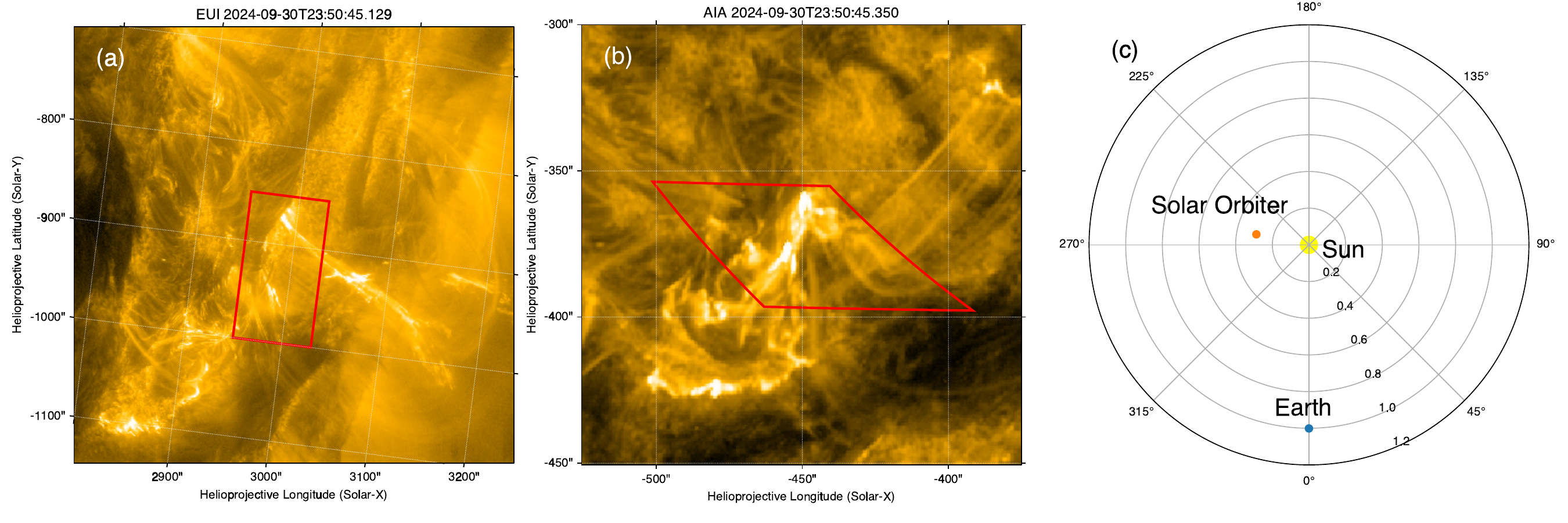}
    \caption{Snapshots of the erupting filament and location of the Solar Orbiter. (a) EUI/HRI 174 \AA\, observation, with time calibrated to Earth time. (b) AIA 171 \AA\, observation. The red boxes in (a) and (b) indicate the same region. (c) Positions of Solar Orbiter and Earth at 23:50 UT on September 30th, 2024.}
    \label{fig:1}
\end{figure*}

In this study, we used data from the HRI$_{\rm EUV}$ telescope (High Resolution Imager in the EUV) of the EUI instrument onboard Solar Orbiter, taken on September 30th, 2024 \citep{euidatarelease6}. HRI$_{\rm EUV}$ operated during 56 min at 2s-cadence thereby collecting 1680 images each with 1.65 s exposure time. It images the solar corona in 174 \AA\, passband. The temperature response function of this passband peaks at the temperature of $\log(T/\mathrm{K})=6.0$ \citep{Berghmans2021,chenYJ2021}. During the observation, the Solar Orbiter was at a distance of 0.2926 au from the Sun, resulting in a pixel size of 0.105 Mm. Its position is shown in Figure \ref{fig:1}(c). The cadence of the dataset is 2 s, with an exposure time of 1.65 s.

We focused on AR13842, located at the west limb within the EUI/HRI$_{\rm EUV}$ field of view (see Figure \ref{fig:1}(a)). From 23:39 UT to 23:46 UT, an erupting filament was observed, with significant intensity enhancement and the untwisting of its threads (magnetic field lines). 
During this time, there is also a M7.6 flare at this active region. The energy release mechanism of the flare was studied recently by \cite{Chitta2025}.
The filament eruption was also observed by the Atmospheric Imaging Assembly (AIA; \citealp{Lemen2012}) onboard the Solar Dynamics Observatory (SDO; \citealp{Pesnell2012}). 
To identify the same region in both instruments, we transformed the coordinates along the four edges of the red box in Figure \ref{fig:1}(a) from the EUI/HRI$_{\rm EUV}$ coordinate frame to the AIA coordinate frame using \texttt{SkyCoord.transform\_to} from the \texttt{astropy.coordinates} module in Python. The resulting coordinates were used to draw the red polygon in Figure \ref{fig:1}(b), outlining the same region as observed by AIA.
However, due to the lower resolution of SDO/AIA and the projection effect, nanojet events could not be captured by AIA. Therefore, we only use the Solar Orbiter/EUI data in this study.

\section{Results}\label{sec:res}

\begin{figure*}
    \centering
    \includegraphics[width=1.0\textwidth]{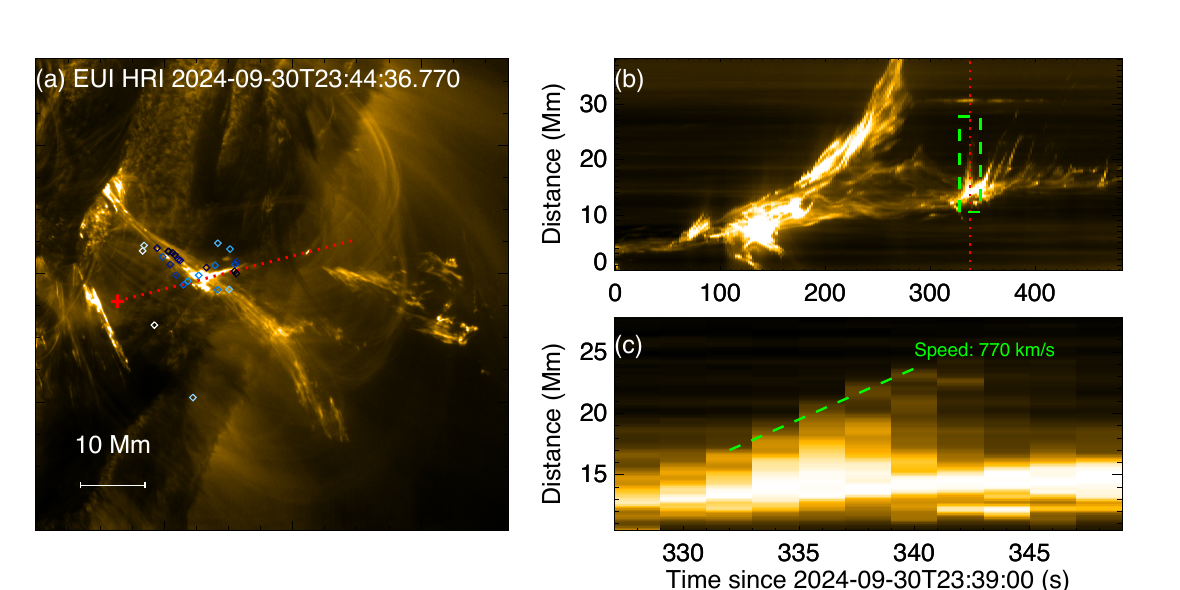}
    \caption{Nanojet No. 13 originated from the untwisting filament. Panel (a) shows the erupting filament and the nanojet, with a red dotted line overplotted along the direction of the nanojet. The locations of all identified nanojet events are marked with diamonds, where darker colors correspond to larger event numbers. (b) Time-distance map generated along the slit. The distance starts at the bottom left of the slit indicated by the red cross in (a). The red dotted line corresponds to the time shown in panel (a). (c) A zoomed-in view of the time-distance map, corresponding to the green dashed box in panel (b). A green dashed line is overplotted along the jet trajectory, with its slope indicating the jet speed of 770 km s$^{-1}$. An animation of this figure is available in the online article.}
    \label{fig:2}
\end{figure*}

During the untwisting process of this erupting filament, we observed 27 nanojet events, all oriented nearly perpendicular to the filament’s spine. In Figure \ref{fig:2}, we present one of the largest events (No. 13), lasting for about 28 s. A time-distance map (panel b) was generated along the ejection direction of the jet, from which we can also observe the untwisting motion of the filament. During this untwisting, it is likely that two magnetic field lines or flux tubes with a sufficient misalignment angle were created, triggering component reconnection. Previous numerical simulations have found that after the reconnection, plasma or plasmoids are ejected from the reconnection site \citep{Antolin2021,Pagano2021}. From the zoomed-in version of the time-distance map shown in Figure \ref{fig:2}(c), the speed ($v$) of the jet is estimated to be 770 km s$^{-1}$. 
To measure the jet's maximum length ($L$), we manually identified the time step at which each jet appears to reach its maximum extent. For this event, $L$ is approximately 6.6 Mm. The jet width ($d$) was measured approximately at the midpoint along its length. We determined the width by fitting the intensity profile perpendicular to the jet axis with a Gaussian function, yielding a value of 0.7 Mm (full width at half maximum). Assuming that the jet is cylindrical in shape, we calculate the volume as $V=\pi L(d/2)^2$ \citep[e.g.,][]{Hou2021}.

To estimate the kinetic energy, we used the formula $E_\mathrm{k} = 0.5n_\mathrm{e}m_\mathrm{p}v^2V$, where $n_\mathrm{e}$ is the electron number density, $m_p$ is the proton mass. Unfortunately, this jet event could not be clearly identified in SDO/AIA data due to interference from background and foreground structures, limiting our ability to obtain density information through methods like differential emission measure. Therefore, we assume a typical coronal electron density $n_\mathrm{e} = 10^9$ cm$^{-3}$, consistent with reported values \citep{Hou2021,Sukarmadji2024}. This assumption leads to an estimated kinetic energy of $1.2\times10^{25}$ erg.
Similarly, by assuming a jet temperature ($T$) of 2 MK \citep{Antolin2021}, we roughly estimated the thermal energy as $E_\mathrm{t}=3n_\mathrm{e}k_\mathrm{B}TV\sim1.9\times 10^{24}$ erg. 

The total magnetic energy released, $E_\mathrm{m}$, can be estimated by adding the kinetic and thermal energies: $E_\mathrm{m} = E_\mathrm{k} + E_\mathrm{t}$, and it is related to the magnetic field strength $B$ through the following equation \citep[e.g.,][]{Priest2014,ChenHD2020}:
\begin{equation}
    E_\mathrm{m}=E_\mathrm{k}+E_\mathrm{t}=\frac{B^2}{8\pi}V\,.
\end{equation}
This gives a magnetic field of approximately 20 G. Given the strong magnetic fields typically found in active regions, this derived value is at least reasonable in order of magnitude.

\begin{figure*}
    \centering
    \includegraphics[width=1.0\textwidth]{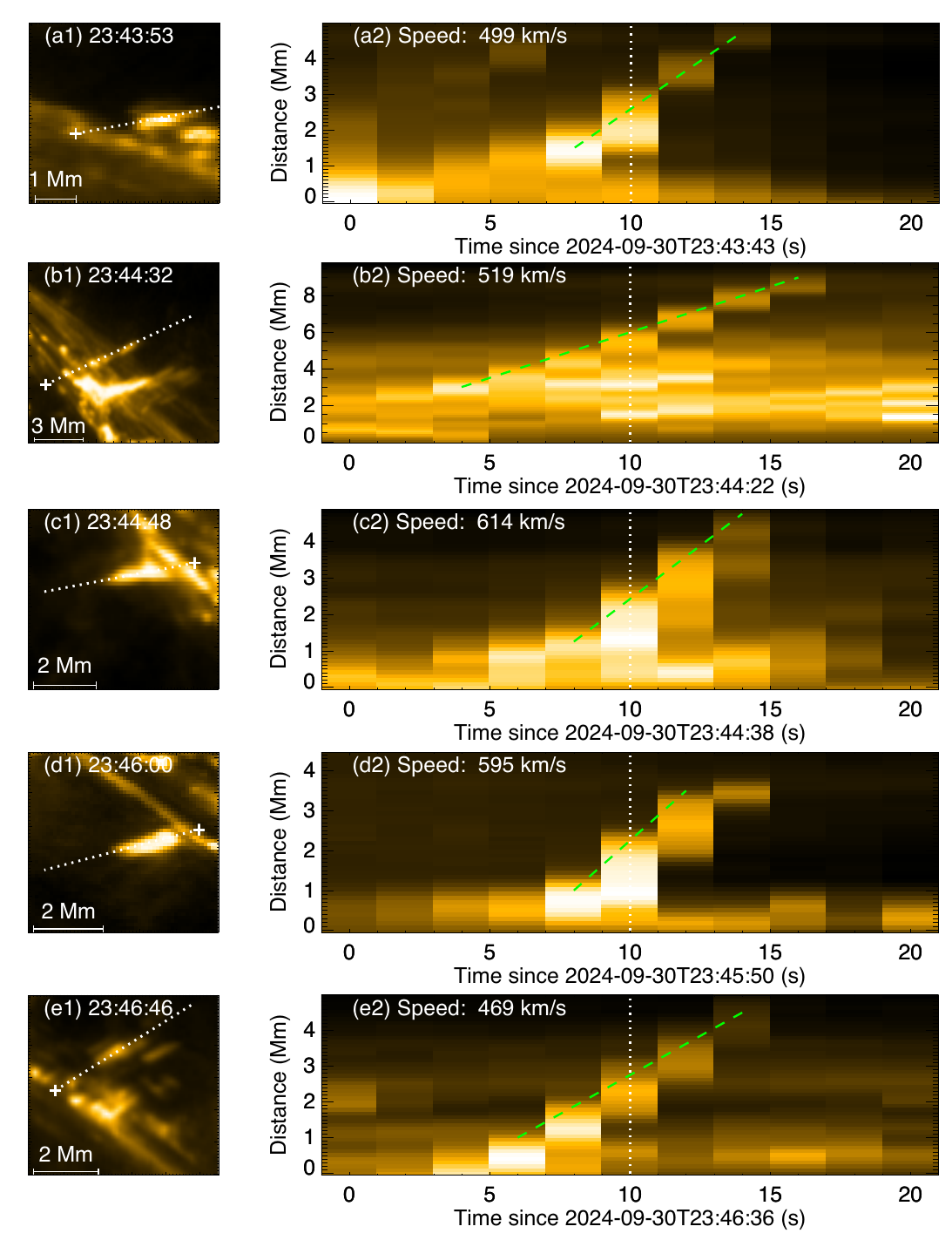}
    \caption{Additional examples of nanojets and their speeds obtained from time-distance maps.  The dotted and dashed lines in the figure are similar to those in Figure \ref{fig:2}. Rows one to five correspond to nanojet No. 6, No. 11, No. 15, No. 17, and No. 27, respectively. An animation of this figure is available in the online article.}
    \label{fig:3}
\end{figure*}

In Table \ref{tab:1}, we summarized the properties of the observed 27 nanojets, including their occurrence time, speed, length, width, kinetic energy, and duration (or lifetime). Five additional examples are presented in Figure \ref{fig:3}. These nanojets are smaller than jet No. 13 shown in Figure \ref{fig:2} but still exhibit large speeds around 500 km s$^{-1}$. 
Some events appear as ejecting blobs, which are possibly plasmoids moving outward from the reconnection sites (see also \citealp{ZhangQM2014,Ni2017,ChenJie2022,Hou2024}).

\begin{figure*}
    \centering
    \includegraphics[width=1.0\textwidth]{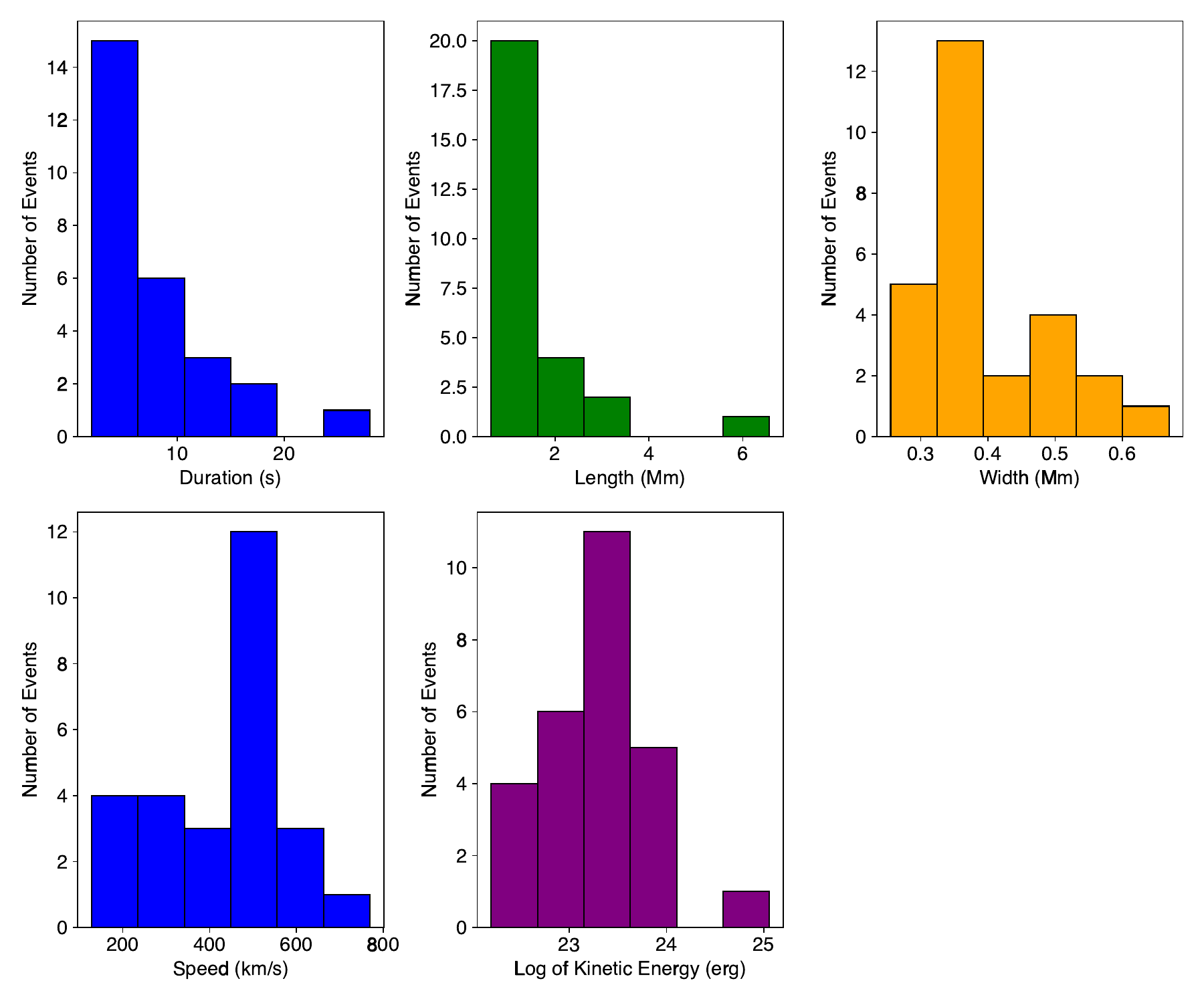}
    \caption{Histograms of nanojet properties including durations, (maximum) length, width, speed, and kinetic energy.}
    \label{fig:4}
\end{figure*}

In Figure \ref{fig:4}, we show histograms of the observed properties. Most events have a duration or lifetime of less than 10 s. Particularly, several appear only in one or two frames (with durations of 2 or 4 s), indicating a highly dynamic nature. For such transient events, their exact durations cannot be accurately determined, and the presented values of 2 and 4 s should be considered some rough estimates. Moreover, for such short-lived events, it is also hard to derive their speeds from time-distance maps. However, by tracking the movement of the jet fronts between adjacent frames, we can still estimate their speeds (similar to \citealp{Chitta2023}). The average length and width of these nanojets are 1.5 Mm and 0.4 Mm, respectively, which are slightly smaller than previously-reported values \citep{ChenHD2017, ChenHD2020, Antolin2021, Sukarmadji2022, Patel2022}. The speeds of these nanojets range from 128 to 770 km s$^{-1}$, with a majority exhibiting speeds around 500 km s$^{-1}$. Regarding kinetic energy, all events fall within the range of $10^{22}-10^{25}$ erg, with an average of $7.2\times 10^{23}$ erg. Thus, they can still be classified as nanoflare-type events, although their energy are slightly less than previously reported due to their smaller sizes.

We note that the measured speeds represent only the plane-of-sky (POS) components of the true speeds. Due to the absence of simultaneous high-resolution observations from multiple viewing angles, it is not possible to reconstruct the full 3D velocity vectors of the nanojets. Therefore, the estimated speeds and kinetic energies should be considered lower limits of their actual values.

The direction of nanojets is also a topic of interest in previous studies. For instance, \citet{Antolin2021} and \citet{Sukarmadji2022} found that most nanojets propagate inward relative to the curvature radius of the hosting coronal loops, with only two cases showing outward motion. In contrast, \citet{Patel2022} reported a higher fraction of outward-propagating events (4 out of 10 cases). In our study, however, it is difficult to define the propagation direction as inward or outward, since the nanojets originate from an erupting filament rather than from curved loop systems. As shown in Figure~\ref{fig:2}(a) and the accompanying movie, the filament extends roughly from the upper left to the lower right of the field of view. The identified nanojets are observed to eject in both directions—toward the lower left or the upper right—as indicated in the final column of Table~\ref{tab:1}. Specifically, 14 out of 27 events propagate toward the upper right, while the remaining 13 move in the opposite direction. This nearly symmetric distribution is consistent with expectations. Previous studies have shown that magnetic field line curvature can induce reconnection asymmetries and thus influence the directional preference of nanojets \citep{Pagano2021}. However, in the case of an erupting filament, the effect of curvature is minimal, leading to no strong directional bias.

\section{Discussion and Conclusion}\label{sec:diss}

In this study, we used high-resolution EUV observations from the Solar Orbiter/EUI to investigate nanojet events in an erupting filament. We detected 27 small-scale jet-like events in a direction roughly perpendicular to the background filament threads (field lines). These events can be classified as nanojets driven by component reconnection for several reasons: 1) their morphology and properties are similar to those reported in previous studies \citep{ChenHD2017,ChenHD2020,Antolin2021,Patel2022,Sukarmadji2022,Sukarmadji2024}; 2) the energy of these events generally aligns with the definition of nanoflares (around $10^{24}$ erg; \citealp{Parker1988}); 3) component reconnection can be easily triggered when magnetic field lines form misalignment angles \citep{Antolin2021,Pagano2021}, and such misalignment should be ubiquitous in an untwisting filament as observed here. Therefore, component reconnection presents a natural and plausible mechanism responsible for the formation of these jets.

A major difference between our observations and previous ones is the significantly higher jet speed observed in this study. Previouly reported speed values of nanojets are mostly less than 300 km s$^{-1}$ \citep{ChenHD2017,ChenHD2020,Antolin2021,Patel2022,Sukarmadji2022,Sukarmadji2024}. While in this work, 22 out of 27 nanojets have speeds greater than 300 km s$^{-1}$, and 16 out of 27 nanojets have speeds exceeding 450 km s$^{-1}$. The highest speed even reaches $\sim$800 km s$^{-1}$, close to the typical coronal Alfv\'{e}n speed. 
Such high speeds are even rarely observed in other types of jets in the upper atmosphere \citep[e.g.,][]{tian2014,Liu2016,LiXH2018,Sterling2020,Hou2021,panesar2023,Musset2024}. \cite{Cirtain2007} reported that X-ray jets in coronal holes have a fast component with a speed of around 800 km s$^{-1}$. However, this component appears intermittently, and the jet is more dominated by a slower component with a speed of around 200 km s$^{-1}$. As for previously identified small-scale jets, such as microjets \citep{Hou2021}, jetlets \citep{Panesar2019}, moving bright dots \citep{tiwari2019,tiwari2022}, and picoflare jets \citep{Chitta2023}, their speeds are all far below 800 km s$^{-1}$. Thus, the jet speed observed here is unprecedented, not only for nanojets driven by the component reconnection driven but also for other types of small-scale jet-like events.

Two factors might explain the high-speed nature of these nanojets. First, the high temporal and spatial resolution of this observation is crucial. A 2-s cadence is shorter than those of AIA (12 s), IRIS (9-18 s), and Hi-C 2.1 (4.4 s). Some nanojet events in our study have a very short lifetime ($\lesssim$ 4 s), making them only detectable with the 2-s cadence of EUI/HRI$_{\rm EUV}$. 
The average lifetime of these nanojets (8 s) is also much shorter than those reported for other small-scale jet-like phenomena \citep[e.g.,][]{tiwari2019,tiwari2022,Panesar2021,panesar2023,Hou2021}. Additionally, we also observed some low-speed nanojets in this work, suggesting that previous nanojet observations may have only captured the low-speed part due to the instrument's selection effect. Furthermore, the nanojets identified in our study are also smaller in size compared to many previously reported small-scale jets \citep{Panesar2021,panesar2023,Hou2021,Antolin2021,Sukarmadji2022,Sukarmadji2024,Patel2022}. It is likely that only with sufficiently high spatial resolution can we detect jets with such large speeds.
Therefore, our findings emphasize the importance of high-cadence and high-resolution observations in future solar missions for unveiling new phenomena and dynamics in the solar atmosphere. 

Second, the unique physical environment of our nanojet events may also favour the occurrence of high speeds. Previously nanojets were often observed in lower-temperature channels, such as 1400 \AA\, and 304 \AA\,, corresponding to the transition region. The nanojets studied by \cite{Antolin2021}, originating from coronal rain strands, have densities one or two orders of magnitude larger than typical coronal values ($\sim10^9$ cm$^{-3}$). In such an environment, the local Alfv\'{e}n speed could be on the order of 100 km s$^{-1}$. However, at the coronal heights above active regions, the Alfv\'{e}n speed can reach up to 1000 km s$^{-1}$ \citep[e.g.,][]{Anfinogentov2019,Anfinogentov2019radio}, which is comparable to the highest speed observed in this study. It is also worth noting that \cite{Patel2022} reported nanojets in the corona of active regions but did not observe speeds greater than 250 km s$^{-1}$. Their nanojets were found in relatively stable loops, whereas the nanojets in our study are produced during a filament untwisting process. The untwisting motion likely facilitates the formation of larger misalignment angles between adjacent magnetic field lines, leading to more efficient energy release and higher jet speeds. 

We note that, due to the lack of further information such as the magnetic field strength and configuration, currently we can only provide aforementioned qualitative explanations for the observed high speeds. Our next step will be to investigate these processes further using numerical simulations.

The high speed of these nanojets provides new insights into the nanoflare heating process. Our findings emphasize the importance of high-resolution EUV imaging in studying these phenomena, and provide evidence that component reconnection can produce ejecting plasma (plasmoids) at speeds up to 800 km s$^{-1}$. Larger speeds generally imply that a larger portion of magnetic energy is released, and therefore suggest a greater energy dissipation rate, which is significant for the nanoflare heating mechanism. Future studies using similar observations will also help validate various nanoflare heating models \citep{Klimchuk2017, Pontin2020}. Additionally, the study of nanojets also enhances our understanding of the magnetic reconnection process in the solar atmosphere, which is central to various solar activities. Furthermore, the discovery of the unprecedented high speeds for these nanojets suggests that the evolution and heating of filaments during solar eruptions may be more complex than previously thought, with fine structures and dynamic processes contributing to the energy release.

The high-speed nature of these nanojets may also influence the estimation of their length and kinetic energy. The length presented in Table \ref{tab:1} is the apparent length which could be overestimated due to the dynamic blurring effect.
For an exposure time of $t_{\rm exp}=1.65$ s, the dynamic blurring length along the jet's ejecting direction can be estimated as $L_{\rm db}=vt_{\rm exp}$. For our observed nanojets, this ranges from 0.2 to 1.2 Mm, potentially leading to a considerable overestimation of jet length. However, it is challenging to precisely quantify the contribution of this blurring to the apparent length. We calculated the ratio $(L-L_{\rm db})/L$ ($L$ taken from Table \ref{tab:1}) and found an average value of 46\%. If the actual jet length (and consequently the kinetic energy which is proportional to the length) is around 46\% of the measured value, the estimated average kinetic energy becomes around $3.6\times10^{23}$ erg. Moreover, while dynamic blurring indeed affects length and energy estimates, it does not significantly influence our key findings regarding the unprecedentedly high speeds of these nanojets. A more detailed investigation into the dynamic blurring effect will be the focus of future work. To avoid dynamic blurring with pixels of 105 km and observed speeds around 500 km s$^{-1}$, one would need exposure times around 0.2 s. The EUI Major Flare Watch campaigns (Ryan et al. 2025, submitted) that started recently, alternate regular HRI$_{\rm EUV}$ images with ones with shorter exposure time (as short as 0.04 s) to image flare kernels. It remains to be analyzed if other flares observed during the EUI Major Flare Watch campaigns show nanojets and if they are bright enough to be identifiable in 0.04 s exposures.

In conclusion, this letter reports the detection of reconnection nanojets in an erupting untwisting filament using high-resolution EUV imaging from Solar Orbiter/EUI. We investigated 27 nanojet events and recorded their properties. The most remarkable finding is that the speeds of these nanojets reach up to 800 km s$^{-1}$, much higher than previously reported. This suggests that these nanoflare-type phenomena can be more dynamic than previously recognized, thus shedding new light on the magnetic reconnection processes in the dynamic corona.

\begin{acknowledgments}

This work is supported by the National Natural Science
Foundation of China (grant No. 12425301 and No. 12073004), the
Strategic Priority Research Program of the Chinese Academy
of Sciences (grant No. XDB0560000), the National Key
R\&D Program of China No. 2021YFA1600500, and the Natural Science Foundation of
Beijing, China (No. 1254055). T.V.D was supported by the C1 grant TRACEspace of Internal Funds KU Leuven and a Senior Research Project (G088021N) of the FWO Vlaanderen. Furthermore, T.V.D received financial
support from the Flemish Government under the long-term
structural Methusalem funding program, project SOUL: Stellar
evolution in full glory, grant METH/24/012 at KU Leuven.
The research that led to these results was subsidized by the
Belgian Federal Science Policy Office through the contract
B2/223/P1/CLOSE-UP. It is also part of the DynaSun project
and has thus received funding under the Horizon Europe
program of the European Union under grant agreement (No.
101131534). Views and opinions expressed are however those
of the author(s) only and do not necessarily reflect those of the
European Union and therefore the European Union cannot be
held responsible for them. 
H.C.C. was supported by the Yunnan Provincial Basic Research Project (202401CF070165).
D.Y.D. was supported by Beijing Natural Science Foundation (1244053), NSFC grant 12403065. 
H.T. also acknowledges support from the New Cornerstone Science Foundation through the Xplorer Prize.
Solar Orbiter is a space mission of international collaboration between ESA and NASA, operated by ESA. The EUI instrument was built by CSL, IAS, MPS, MSSL/UCL, PMOD/WRC, ROB, LCF/IO with funding from the Belgian Federal Science Policy Office (BELSPO/PRODEX PEA 4000112292 and 4000134088); the Centre National d’Etudes Spatiales (CNES); the UK Space Agency (UKSA); the Bundesministerium für Wirtschaft und Energie (BMWi) through the Deutsches Zentrum für Luft- und Raumfahrt (DLR); and the Swiss Space Office (SSO).
We also acknowledge support by ISSI and ISSI-BJ to the team ``Small-scale Eruptions on the Sun".
\end{acknowledgments}

\vspace{5mm}
% \facilities{HST(STIS), Swift(XRT and UVOT), AAVSO, CTIO:1.3m,


% \software{astropy \citep{2013A&A...558A..33A,2018AJ....156..123A},  
%           Cloudy \citep{2013RMxAA..49..137F}, 
%           Source Extractor \citep{1996A&AS..117..393B}
%           }



% \appendix

\bibliography{sample631}{}
\bibliographystyle{aasjournal}



\end{document}